\documentclass[preprint,12pt]{elsarticle}

\usepackage{amssymb}

\usepackage{listings}
\usepackage{color}
\usepackage{tikz}
\usepackage{booktabs}
\usepackage{multirow,bigstrut}
\usepackage{tabu}
\usepackage{comment}
\usepackage{todonotes}
\usepackage{color,soul}
\usepackage{url}
\usepackage{hyperref}
\usepackage{pdfcomment}

\lstdefinestyle{customasm}{
    belowcaptionskip=1\baselineskip,
    frame=single, 
    frameround=tttt,
    xleftmargin=\parindent,
    language=[x86masm]Assembler,
    basicstyle=\footnotesize\ttfamily,
    commentstyle=\itshape\color{green!60!black},
    keywordstyle=\color{blue!80!black},
    identifierstyle=\color{red!80!black},
    tabsize=4,
    numbers=left,
    numbersep=8pt,
    stepnumber=1,
    numberstyle=\tiny\color{gray}, 
    columns = fullflexible,
}

\newlength\MAX  
\newcommand*\Chart[1]{#1~\rlap{\textcolor{black!20}{\rule{\MAX}{2ex}}}\rule{#1\MAX}{2ex}}
\newcommand{\approach}[1]{\emph{ACCA}}

\journal{Journal of Systems and Software}

\begin{document}

\begin{frontmatter}



\title{Automating the Correctness Assessment of AI-generated Code for Security Contexts}


\author{Domenico Cotroneo}
\ead{cotroneo@unina.it}
\author{Alessio Foggia}
\ead{alessio.foggia@unina.it}
\author{Cristina Improta}
\ead{cristina.improta@unina.it}
\author{Pietro Liguori\corref{cor}}
\ead{pietro.liguori@unina.it}
\author{Roberto Natella}
\ead{roberto.natella@unina.it}

\affiliation{organization={University of Naples Federico II},
            city={Naples},
            country={Italy}}

\cortext[cor]{Corresponding author}

\begin{abstract}
Evaluating the correctness of code generated by AI is a challenging open problem. 
In this paper, we propose a fully automated method, named \approach{}, to evaluate the correctness of AI-generated code for security purposes. The method uses symbolic execution to assess whether the AI-generated code behaves as a reference implementation.
We use \approach{} to assess four state-of-the-art models trained to generate security-oriented assembly code and compare the results of the evaluation with different baseline solutions, including output similarity metrics, widely used in the field, and the well-known ChatGPT, the AI-powered language model developed by OpenAI.

Our experiments show that our method outperforms the baseline solutions and assesses the correctness of the AI-generated code similar to the human-based evaluation, which is considered the ground truth for the assessment in the field.
Moreover, \approach{} has a very strong correlation with the human evaluation (Pearson's correlation coefficient $r=0.84$ on average).
Finally, since it is a fully automated solution that does not require any human intervention, the proposed method performs the assessment of every code snippet in $\sim 0.17$s on average, which is definitely lower than the average time required by human analysts to manually inspect the code, based on our experience.
\end{abstract}



\end{frontmatter}


\section{Introduction}
\label{sec:introduction}

Artificial Intelligence (AI) code generators use Neural Machine Translation (NMT) models to turn natural language (NL) descriptions into programming code. They represent a powerful asset in the arsenal of cybersecurity professionals and malicious programmers.  Indeed, AI (\textit{offensive}) code generators are becoming an attractive solution to creating \textit{proof-of-concept} exploits in order to assess the exploitability and severity of software vulnerabilities~\cite{yang2023exploitgen,liguori2021evil,ruan2023prompt}, letting the AI helping developers to generate low-level (i.e., assembly) and complex code, with a reduced effort and improved effectiveness.

Despite the dramatic increase in the adoption of AI code generators, they still have limitations and potential drawbacks. For example, they may not always generate code that is \textit{correct}, i.e., code that performs what is required from the NL description, as they may struggle with more complex programming tasks that require human creativity and problem-solving skills, or may incorrectly interpret developers' descriptions.
Furthermore, AI code generators can introduce security vulnerabilities if not properly tested and validated~\cite{DBLP:conf/sp/PearceA0DK22,siddiq2022empirical,llmseceval}. 
For these reasons, assessing the correctness of AI-generated code becomes a crucial challenge. 

From the existing literature, it comes out that one of the most effective ways to assess the correctness of generated code is to perform a manual code review (i.e., \textit{human evaluation})~\cite{DBLP:conf/emnlp/ParvezACRC21,kononenko2016code}. This involves having a human expert review the code and identify any errors or inconsistencies with the NL description.
However, human evaluation has several limitations. First, manual analysis can be a time-consuming process. Indeed, reviewers must carefully examine each line of code and thoroughly test the software to ensure that it meets the intended requirements and NL specifications. This process also requires reviewers to be highly knowledgeable about the programming language, development environment, and intended functionality of the code to provide accurate assessments. 
Moreover, the analysis can be subjective, as different reviewers may have different interpretations of the code and its intended functionality, depending on the expertise and experience of the reviewer. This can lead to inconsistent assessments of code correctness. 
Last but not least, manual analysis is prone to human error, as reviewers may miss subtle errors or inconsistencies in the code, or may introduce errors and biases into their assessments due to factors such as fatigue, distractions, or subjective opinions.
From the above considerations, it is clear that what we gained from the help of AI, we lost due to the manual review.


Unfortunately, currently no fully automated solution can perform the semantic assessment of AI-generated code comparably to human evaluation.
In fact, although existing automated testing and code analysis tools can effectively identify errors or inconsistencies in code, they do not provide any insights into whether the code is what is actually required by developers~\cite{ayewah2008using,bessey2010few,liu2019avatar,pylint}. Moreover, these solutions often require in inputs entire, compilable programs (e.g., entire functions) rather than single code snippets, which is instead often the case with AI-generated code.
Indeed, code snippets produced by models are often not complete programs but rather fragments or components of code that, on their own, might not be directly compilable. This characteristic is due to the nature of the training data available and the current capabilities of AI code generation models, which are adept at generating specific functional code snippets rather than entire, complex applications. This highlights the limitation in evaluating the correctness and functionality of these snippets using traditional software testing methods (e.g., unit-level testing)~\cite{araujo2020far,lukasczyk2023empirical}.

Besides the automated solution issue, there is a more important one, i.e., how to evaluate the correctness of AI-generated code. Indeed, previous studies proposed a large number of \textit{output similarity} metrics, i.e., metrics computed by comparing the textual similarity of generated code with a ground-truth reference implementation~\cite{reiter2009investigation,shterionov2018human,LIGUORI2023120073}. The major advantage of the proposed metrics is that they are reproducible, easily tuned, and time-saving. 
However, in the context of programming code generation, existing metrics are not able to fully reflect the correctness of the code. 

As illustrated in the next section, generated code can be different from the reference but still be correct (e.g., the assembly conditional jumps \texttt{jz} and \texttt{je} are different instructions that can be used to perform the same operation); or, there can be subtle differences between the generated and the reference code, which can be similar yet produce different outputs (e.g., the assembly conditional jumps \texttt{je} and \texttt{jne} are syntactically similar instructions, but they perform the opposite operation).
Hence, it is crucial to develop novel, more accurate methods for automatically evaluating the correctness of AI-generated code. 

This paper proposes a method, named \approach{} (\textit{Assembly Code Correctness Assessment}), to automatically assess the correctness of assembly AI-generated code without any human effort. More precisely, our solution leverages \textit{symbolic execution}, i.e., a state-of-the-art solution for program analysis based on abstract execution, which consists of simulating the execution of a program providing symbolic values to evaluate its behavior, to assess whether the generated code behaves as a reference implementation, despite syntactic differences between the reference and the generated code.

We apply \approach{} to assess four state-of-the-art NMT models in the generation of security-oriented code in assembly language starting from NL descriptions in the English language and compare the results of \approach{} with the human evaluation and several baseline assessment solutions, including a wide range of output similarity metrics and the well-known ChatGPT by OpenAI.
We show that the proposed method provides an almost perfect assessment of the code's correctness and has a very strong correlation with the human evaluation, outperforming all the baseline assessment solutions.

In the following,
Section~\ref{sec:motivation} introduces a motivating example;
Section~\ref{sec:methodology} describes \approach{};
Section~\ref{sec:setup} presents the experimental setup;
Section~\ref{sec:results} shows the experimental results;
Section~\ref{sec:related} presents the related work; 
Section~\ref{sec:conclusion} concludes the paper.

\section{Motivating Example}
\label{sec:motivation}

In code generation tasks, datasets play a crucial role in training and evaluating AI models. These datasets typically consist of pairs of inputs and outputs, where the input is a natural language (NL) description of a coding task, and the output is the code snippet that fulfills the described task. NL descriptions refer to human-readable explanations or specifications of what a particular piece of code is intended to achieve. These descriptions serve a similar purpose to code comments but are used as input for AI code generation models to produce corresponding code snippets.

These corpora are commonly split into \textit{training data}, i.e., the data used to feed the model, \textit{validation data}, i.e., the data used to tune the model's parameters, and \textit{test data}, i.e., the data used to evaluate the model in the generation of the code starting from new NL descriptions (i.e., the NL intents in the test data are never seen by the model in the train and validation data).

The most practical solution to assess the performance of the NMT models in the code generation is to compare, for every NL description of the test data (i.e., the input), the model's prediction with the code snippet (i.e., the output) in the test set, which is considered the \textit{ground-truth} for the evaluation. To this aim, state-of-the-art provides a set of metrics that estimate the similarity between the code generated by NMT models and the code snippets in the test set.
However, output similarity metrics cannot properly assess whether two pieces of code are different but semantically equivalent, i.e., they provide the same output and/or effects although they use different operations (e.g., \texttt{jz label} and \texttt{je label} are different assembly instructions performing the same conditional jump).

State-of-the-art also proposed solutions that apply contrastive learning to discern between semantically similar and dissimilar code samples~\cite{wu2022contrastive,massarelli2021function,ullah2021bindiff}. Code embedding models via contrastive learning are designed to learn rich, high-dimensional representations of code snippets by embedding them into a vector space. While they present an advanced method for capturing the semantic features of code snippets, several challenges limit their immediate applicability as a primary evaluation solution for AI-generated code. First, they require significant computational resources not only for training but also for inference. Evaluating large datasets or conducting extensive experiments can become computationally expensive and time-consuming, limiting the scalability of this solution. Moreover, the effectiveness of code embedding models is heavily dependent on the quality and diversity of the training data. Models trained on limited or biased datasets may not accurately capture the full spectrum of semantic equivalences, potentially skewing the evaluation results. This issue is further exacerbated in our case study, where there is a lack of corpora used for offensive code security, especially in low-level languages. On the contrary, output similarity metrics, which estimate the textual similarity of the model’s predictions with respect to a ground truth reference, are easy to use, time-saving, and require low resource requirements, and represent the most common solution to assess models in the SOTA.

For the aforementioned reasons, \textit{human evaluation} is considered the golden standard for assessing the correctness of the code generated by the models~\cite{evtikhiev2023out}. Through manual inspection of every model's predictions, human evaluators assess if the code generated by the models is \textit{semantically correct}, i.e., if the output is the exact translation of the NL intent into the target programming language. Semantic correctness implies \textit{syntax correctness}, i.e., a code prediction that performs what is described in the NL intent must also adhere to the syntax rules of the target programming languages.
Human evaluation classifies the code as correct or incorrect by assigning a value equal to $1$ or $0$, respectively.



As a simple example, consider the intent ``\textit{transfer EAX contents into EDX register}", which translates, on the 32-bit version of the x86 instruction set architecture (IA-32), to the assembly snippet:

\begin{center}
    \texttt{mov EDX, EAX}
\end{center}

An alternative method to copy the contents of a register into another is by pushing and popping its value onto the stack. Therefore, a semantically equivalent implementation of this copy is the code:

\begin{center}
\begin{tabular}{l}
     \texttt{push EAX} \\ 
     \texttt{pop EDX}
\end{tabular}
\end{center}

Despite the model's prediction being both syntactically and semantically correct, output similarity metrics are not able to grasp the equivalence between the two snippets since they base their calculation on character and/or token similarity. Therefore, this translation results in low scores\footnote{Scores of output similarity metrics range between $0$ and $1$.} for several output similarity metrics widely used in the field (see \S{}~\ref{subsec:metrics}), such as \textit{BLEU-4} ($0.11$) and \textit{Edit Distance} ($0.31$).

The opposite occurs with the intent ``\textit{clear the EDX register and move 5 in the lowest byte of the register}'', which translates to the assembly snippet:

\begin{center}
\begin{tabular}{l}
    \texttt{xor EDX, EDX} \\ 
    \texttt{mov DL, 5}
\end{tabular}
\end{center}

If the model generates the snippet: 

\begin{center}
\begin{tabular}{l}
    \texttt{xor EDX, EDX} \\
    \texttt{mov BL, 5}
\end{tabular}
\end{center}

then prediction and reference differ by a single character, yet the code does not accomplish the same task. Indeed, the lowest byte of \texttt{EDX} is stored in the \texttt{DL} register, while \texttt{BL} contains the lowest byte of \texttt{EBX}.
Automatic metrics fail to account for situations like this. For instance, the \textit{Edit Distance} between these two pieces of code is $0.96$, while the \textit{BLEU-4} is $0.65$, which are considered high values. Differently, a human evaluator would appropriately classify this snippet as semantically incorrect, since it does not perform the intended operation, although it properly respects the syntax of the assembly language.  

However, since the human analyst needs to check the syntax and the semantics of every output generated by the models, human evaluation is often unfeasible. Indeed, the huge amount of data to scrutinize makes the analysis time-consuming and prone to errors. 






\section{Proposed Method}
\label{sec:methodology}

\begin{figure*}[t]
    \centering
    \includegraphics[width=1\columnwidth]{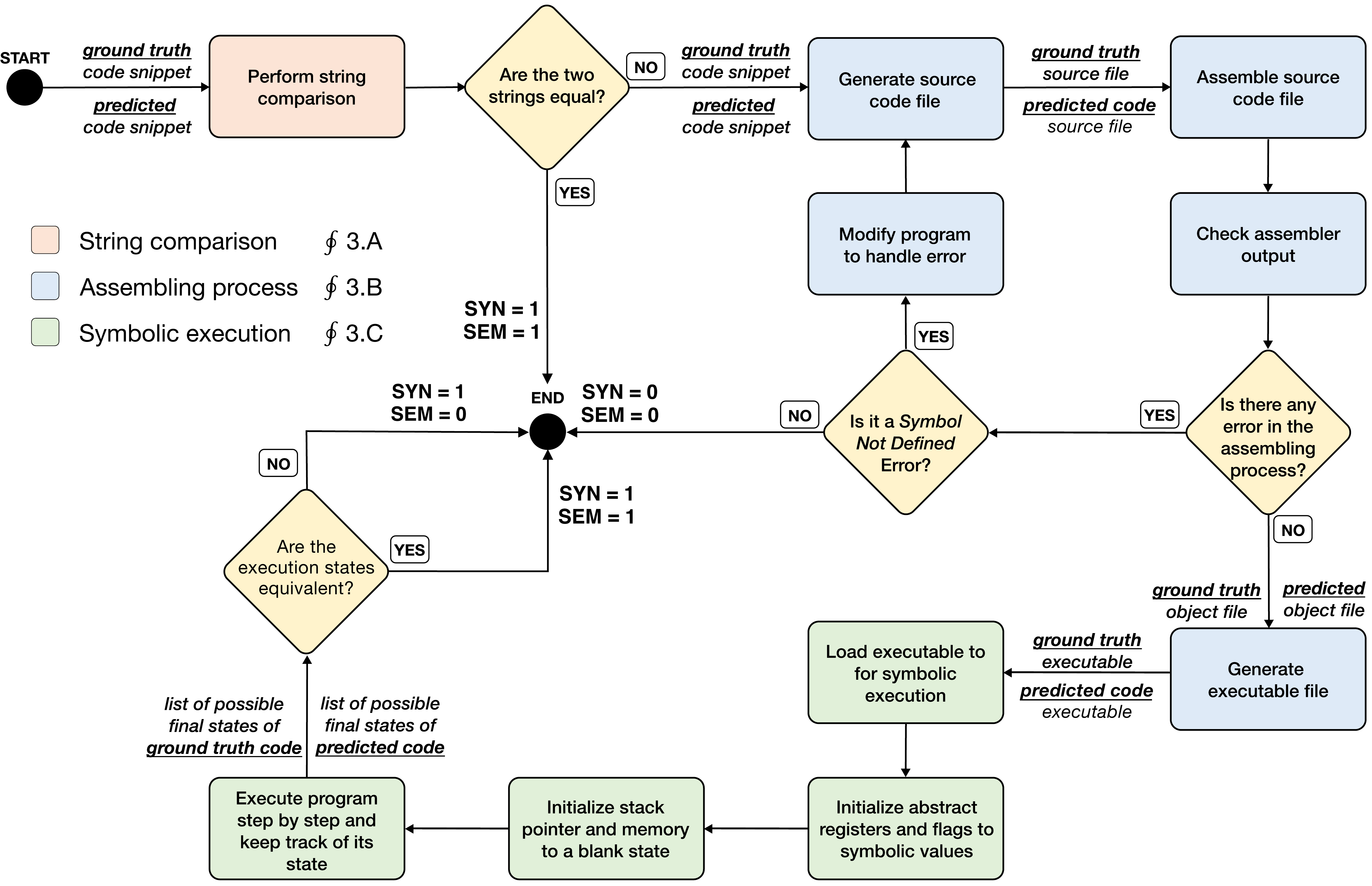}
    \caption{Detailed flowchart of \approach{}.}
    \label{fig:exe_generation}
\end{figure*}

To overcome the limitations in the assessment of AI-generated assembly code described in the previous section, we propose \approach{}, a method to automatically assemble and symbolically execute both the reference (i.e., the ground truth) and predicted snippets (i.e., the code generated by the models). Through symbolic execution, the method simulates the execution of both programs and determines whether, starting from the same state, they terminate producing equivalent results and effects. 

First, \approach{} compares, at the string level, the prediction with the reference to assess its correctness, because, if the prediction is equal to the reference, then we assume the code prediction is correct. 
If this is not the case, the method assesses whether the code is syntactically correct. Indeed, if a code prediction is not structured according to the rules of the target programming languages, it is classified as incorrect.

If this is not the case, i.e., if the prediction differs from the ground truth reference and the prediction is syntactically correct, \approach{} generates two source code files, one for the reference and one for the predicted snippet. It then assembles and links them to produce the executable files needed for the symbolic execution. At this point, the method symbolically executes both files resulting from the assembling process to assess whether they are equivalent.

Finally, \approach{} returns the pair of syntactic correctness value (SYN) and the semantic correctness value (SEM) of the code predicted by the model, which is equal to $1$ when the correctness is verified, $0$ otherwise.
\figurename{}~\ref{fig:exe_generation} shows a detailed flowchart of the syntactic and semantics correctness evaluation process.

In the rest of this section, we detail the phases of the workflow. For the sake of simplicity, we showcase examples of assembly code for the Linux OS running on IA-32 architecture to describe the evaluation process, although the proposed method is not restricted to this specific context, and it can be applied to different instruction set architectures and operating systems.

\subsection{String Comparison}
\approach{} first checks whether the predicted code snippet perfectly matches the ground truth by performing a string comparison. If they match (i.e., the code generated by the model is equal to ground truth), the prediction is considered both syntactically and semantically correct (i.e., SYN=1, SEM=1) and the evaluation ends. Otherwise, i.e., if they differ, the method proceeds with the evaluation of the syntactic and semantic correctness.
The preliminary string comparison is done to speed up the evaluation process by skipping the symbolic execution process when not needed, i.e., when the prediction perfectly matches the reference snippet and is, therefore, correct. 

\subsection{Assembling Process} 
\label{assembling_process}
The purpose of the assembling process is to assess whether each code snippet generated by the model adheres to the syntax rules of the programming language it is written in, i.e., to check whether it is compilable. 
Since NMT is still far from producing entire complex programs, the output predicted by the models is a portion of an entire program (i.e., a single-line or multi-line statement). Thus, \approach{} constructs a complete program by adding the necessary entry point and exit point code. 
For instance, consider the following code snippets for IA-32: 

\begin{center}
\small
\begin{tabular}{l}
     \texttt{cmp EAX, EBX} \\ 
     \texttt{je loop} \\
     \texttt{pop ECX}
\end{tabular}
\end{center}

This code compares the contents of two registers and, based on the result, either performs a jump operation or reads from the stack and saves the value into another register.
The snippet is syntactically correct according to the assembly language, yet it is incomplete for the execution since code snippets need to contain a minimal set of information to be properly executed.

In the case of Linux OS, these instructions are kept in the \texttt{text} section, which must begin with the declaration \texttt{global \_start} to define the initial program entry point. 
Moreover, to make sure the program terminates correctly, the proposed method also includes a fictitious label that represents the exit address the code jumps to at the end of its execution. This label is declared in the \texttt{data} section, which contains initialized read-write data.

Therefore, to have a complete program, the code is modified as follows:

\begin{center}
\small
\begin{tabular}{l}
     \texttt{section .data} \\
     \texttt{exit\_addr db 0x56} \\
     \texttt{section .text} \\
     \texttt{global \_start} \\
     \texttt{\_start:} \\
     \hspace{1cm} \texttt{cmp EAX, EBX} \\ 
     \hspace{1cm} \texttt{je loop} \\
     \hspace{1cm} \texttt{pop ECX}  \\  
     \hspace{1cm} \texttt{jmp exit\_addr} 
\end{tabular}
\end{center}


Once the whole program is created, \approach{} generates a source file and leverages an assembler to assess its syntactic correctness. Indeed, if the programs compile, then all the instructions of the programs are syntactically correct, therefore also the code generated by the model respects the structure of assembly programming language. 
There are three possible output scenarios for the compilation:
\begin{itemize}
    \item \textit{No errors}, in which the assembling process is completely correct;
    \item \textit{Warnings}, in which the assembler reports some type of warning (e.g., word data exceeds bounds, invalid register size specification ignored, etc.), but the compilation still terminates without errors;
    \item \textit{Errors}, in which the assembling process results in an error that prevents the code from being assembled.
\end{itemize}

In the first two cases, the output produced by the model is considered syntactically correct (i.e., SYN=1). 
Warnings are considered acceptable since they indicate issues involving bad practice, but are not severe enough to prevent the code compilation.
Indeed, compiler warnings are raised for potential issues that do not violate the language's syntax rules (e.g., uninitialized space declared in .text section, invalid register size specification ignored). Moreover, these warnings are mainly due to the lines of code added by our method to perform the assessment. Therefore, distinguishing between code that compiles with warnings and code that compiles without warnings introduces a level of granularity that goes beyond the scope of syntax rule adherence as their presence does not mean that the code is syntactically incorrect.

When the compilation produces an error (i.e., the third case), we investigate the nature of the error. More precisely, we focus on a specific category of the error raised by the compiler, the \textit{Symbol-Not-Defined (SND) errors}, which occur when the code contains a symbol (e.g., a label or variable) that has not been previously defined or initialized.
We handle this category of errors appropriately. Indeed, since the predicted snippets contain only one or a few instructions, they might reference a label or variable defined in a different portion of the program, which leads to an assembling error even when the program is syntactically correct.
Indeed, in the context of AI-generated code, it is common to deal with incomplete code instances or to assume context that is not explicitly provided within the snippet itself. This is particularly true for assembly language, where labels and jump instructions play a critical role in control flow. Our decision to specifically manage SND errors stems from their common occurrence in AI-generated assembly code and the nature of the code generation tasks we are evaluating. Unlike other compilation errors that might indicate fundamental syntactic misunderstandings by the AI (e.g., incorrect instruction syntax), SND errors can often be attributed to the context-dependent and fragmentary nature of the generated code snippets. Therefore, by addressing SND errors, the tool can evaluate the syntactic correctness of a code snippet to mitigate the incomplete context typically associated with AI-generated code.

For instance, consider, again, the previous code snippet: the first instruction compares the contents of the \texttt{EAX} and \texttt{EBX} registers and, if they are equal, the execution jumps to the \texttt{loop} label. However, this symbol has not been defined yet. To handle these cases, we analyze the assembler output to determine the missing symbol's name and include it in the source code file as a fictitious label. This label simply points to a jump operation to the previously defined exit address (i.e., \texttt{myExitAddr}). Indeed, the destination of the jump is not significant for the evaluation since we are only interested in checking the correctness of the instructions generated by the model.
Therefore, after a SND error, \approach{} further modifies the program as follows:

\begin{center}
\small
\begin{tabular}{l}
     \texttt{section .data} \\
     \texttt{myExitAddr db 0x56} \\
     \texttt{section .text} \\
     \texttt{global \_start} \\
     \texttt{\_start:} \\
     \hspace{1cm} \texttt{cmp EAX, EBX} \\ 
     \hspace{1cm} \texttt{je loop} \\
     \hspace{1cm} \texttt{pop ECX}  \\ 
     \hspace{1cm} \texttt{jmp myExitAddr} \\ 
     \texttt{loop:} \\
     \hspace{1cm} \texttt{jmp myExitAddr}
\end{tabular}
\end{center}

Once the source code file is modified accordingly, \approach{} repeats the assembling process as before to check if the compilation ends with no errors or warnings. In this case, \approach{} assigns the SYN score equal to 1 and continues the evaluation to check the semantic correctness.

When the compilation ends with errors different from the SND, such as an invalid combination of opcode and operands, expression syntax error, etc., then \approach{} labels the model's prediction as syntactically incorrect (SYN=0). Since a snippet syntactically incorrect is also semantically incorrect, then the evaluation process terminates, assigning the SEM score equal to zero.
A source code file is generated and assembled for both the ground truth and the predicted code snippet.

\subsection{Symbolic Execution}

To evaluate the semantic correctness, \approach{} leverages the symbolic execution. To this aim, the method needs the program executable. If the assembling process ends correctly, the assembler outputs an object file, which is then fed to the linker to complete the process. 

Since the same operation may be correctly implemented in different ways, a simple textual comparison with the reference is not enough to assess the semantic correctness of a program. 
We still consider the ground truth as the reference for the correct implementation of the intended code. However, we do not limit the comparison to a textual similarity, but we examine the \textit{system state} at the end of the execution of both the reference and the generated code. Indeed, two programs that implement the same functionality using different operations can be considered semantically equivalent if they result in the same final system state.
Since the final execution state depends on the inputs and initial state of the program, we need to compare the state produced by both programs for every possible combination of inputs and initial state. 

Symbolic execution is a state-of-the-art solution for program analysis based on abstract execution. 
It consists in simulating the execution of a program providing symbolic values as its input instead of concrete ones. The result of this execution is a set of output values expressed as functions of the input symbols. 
\approach{} uses symbolic execution to determine all the existing execution paths and all possible corresponding output system states. It then compares the set of output system states of the generated program with the set of output system states of the ground truth program: if they match, then the programs are semantically equivalent (SEM=1), otherwise, the method classifies the model's prediction and the ground-truth as not semantically equivalent (SEM=0).

\begin{figure}[t]
    \centering
    \includegraphics[width=0.75\columnwidth]{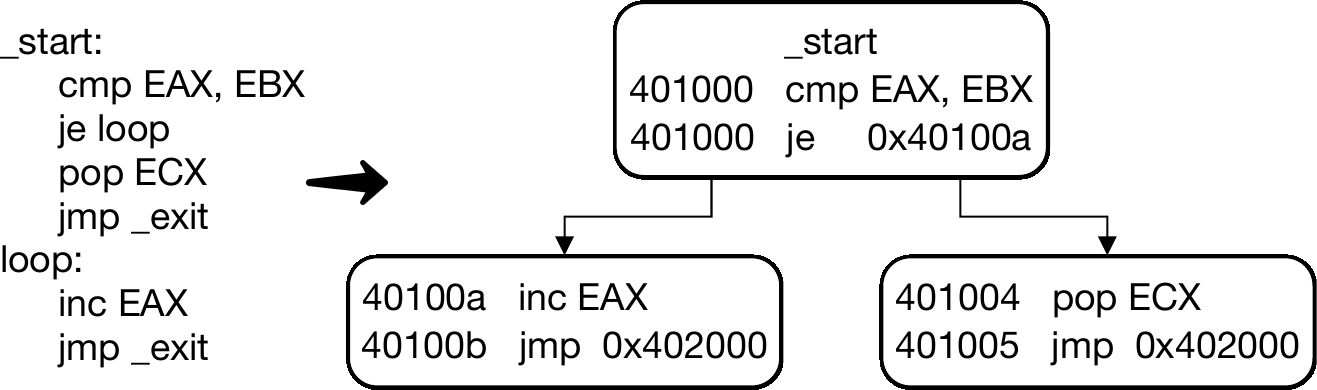}
    \caption{Example of a code snippet represented as a sequence of basic blocks.}
    \label{fig:basic_blocks}
\end{figure}

To symbolically execute the programs, we use a \textit{binary analysis platform} (BAP) that loads each executable and provides a complete abstract representation of the target architecture, CPU registers, memory address space, and stack. The program is conceived as a sequence of \textit{basic blocks}, i.e., a straight-line code sequence with no branches, and the interconnections between the blocks represent the jump operations. An example is shown in Fig. \ref{fig:basic_blocks}: the program compares the contents of two registers and, if they are equal, the execution jumps to a specific address, otherwise, it jumps to the next instruction. Each possible branch is the entry point of a different basic block, which contains a sequence of operations that can be executed in order and one last instruction that causes the execution to move to another basic block.

\approach{} begins the execution by initializing the abstract registers and flags to the same symbolic values; the method also sets the value of the stack pointer and initializes the memory to a blank state.
Then, it executes the program by simulating each instruction step by step and keeping track of its state at each given step. 
During the execution of the programs, performing operations, such as arithmetic operations, comparisons, assignments, etc., that involve a variable (e.g., \texttt{if X > 10}) will yield an \textit{execution tree}, i.e., a tree of operations formed by all the possible paths of the program, which encode all branch decisions taken up to a specific point.
Execution trees are then translated into constraints, i.e., formulas that express a set of assumptions on the symbolic outputs of the execution. These constraints are finally solved by a \textit{satisfiability modulo theories} (SMT) solver, which is typically employed in program verification to verify whether constraints can be satisfied by some assignment of concrete values to the program’s symbolic arguments~\cite{baldoni2018survey}.
As an example, consider the following simple constraints:

\begin{center}
\small
\begin{tabular}{r}
     \texttt{x > y} \\
     \texttt{y > 2} \\
     \texttt{10 > x} \\
\end{tabular}
\end{center}

The SMT solver treats them as assertions that must be satisfied with the valid values of symbolic variables. It, therefore, outputs a value that is consistent with all the constraints (e.g., x=4, y=3).

\approach{} symbolically executes both the reference code and the predicted snippet: assuming that the program state at the beginning of the execution is identical, if the programs are semantically equivalent, then their state is also identical at the end of the execution. Therefore, to assess the semantic correctness of the generated code compared to the ground truth, the proposed method checks whether the states of the architecture are equal at the end of both executions. The program state includes:

\begin{itemize}
    \item \textit{state of the registers}, i.e., the contents of the abstract CPU registers;
    \item \textit{state of the flags}, i.e., the  abstract status register that keeps track of the CPU state, including the result of arithmetic operations (e.g., carry flag, overflow flag, etc.) and interruptions;
    \item \textit{values on stack}, i.e., the contents of the memory area used to store temporary values;
    \item \textit{path constraints}, i.e., the condition of input values, defined over the previous items, that leads to the corresponding final state.
\end{itemize}

Discerning between register states, flag states, stack values, and path constraints is integral in our method to accurately assess the state of execution and, by extension, the semantic equivalence of code snippets. The state of execution of a program at any point includes not just the data it operates on (registers and stack values) but also the conditions under which different operations are executed (flags and path constraints). Hence, each component provides unique insights into how the program behaves and interacts with the system's resources. For instance, register states give information about the current data being processed, flag states indicate the outcomes of previous operations (e.g., whether the last arithmetic operation resulted in a zero or a carry), stack values reveal the function call and return addresses along with local variables, and path constraints help understand the decision-making process within the program. Together, these components offer a complete picture of the program's execution path and its interactions with the system's memory and CPU. 
Moreover, each component's state is interrelated and influences the others during execution. For example, the outcome of a conditional branch (determined by flag states) will affect which path the program takes, subsequently influencing register and stack states. Assessing these components in isolation would lose the context of how the program transitions from one state to another, leading to a fragmented understanding of the program's behavior.

\approach{} compares the state of each \textit{leaf} node, i.e., the final \textit{states} at the end of each path in the execution tree representing the program, of both executables.
To compare the leaf nodes, the method constructs a set of lists for every final basic block of the two programs. Each set contains a list of register values, flag values, boolean constraints, and stack values. For example, a reference program whose execution tree ends with two basic blocks (leaf nodes) has two sets of lists. Each set contains all the values that represent the system state for that particular execution path. If the execution tree of the generated program has the same number of leaf nodes, then each list of the two sets is compared with each list of two sets of the reference program. If there is a correspondence between each leaf of the first program and each leaf of the second program, then they are semantically equivalent (SEM=1) and the evaluation process ends. Contrarily, if the leaf nodes of the two program execution trees do not match, then we conclude that the predicted code is not semantically equivalent to the reference snippet (SEM=0) and the process ends. 

Since the total number of states can grow exponentially in the number of branches, one of the main challenges of symbolic execution is the path explosion problem~\cite{baldoni2018survey}. Indeed, keeping track of a large number of pending branches to be explored impacts both the running time and the space requirements of symbolic execution. The primary causes of path explosion are loops and function calls. One common method to deal with this problem is to bind the path exploration to a limited number of iterations. 
To handle programs whose symbolic execution does not terminate, we set a maximum number of execution steps. Since AI-generated code is typically concise and consists of a few assembly instructions, a correct program concludes its execution in a few execution steps. If it runs for more than \texttt{max\_steps}, then the symbolic execution stops, and the generated program is classified as incorrect (SEM=0).

\subsection{Implementation Details}
To assemble the programs and generate the executable files, we rely on the wide set of available open-source software. For the previous examples (on the IA-32), we used the Netwide Assembler (NASM)~\cite{nasm}, an 80x86 and x86-64 assembler that supports a wide range of executable formats, including Linux, BSD, and Windows operating system formats. As a binary analysis platform, we use ANGR~\cite{shoshitaishvili2016sok}. ANGR provides support for a variety of CPU architectures, including ARM, MIPS, PPC, and x86 processors. It comprises a series of sub-components that implement the different steps necessary for the symbolic execution: to disassemble the executables and lift the binary code to an intermediate representation; to simulate the program state and execution, including registers and memory; and to solve the generated constraints, using the z3~\cite{10.5555/1792734.1792766} SMT solver as a backend. We set the maximum number of execution steps \texttt{max\_steps} to 100 to avoid infinite loops.
Our implementation runs on both Linux and Windows OS.
We publicly shared the implementation of \approach{} on GitHub~\footnote{\url{https://github.com/dessertlab/ACCA}}.

\section{Experimental Setup}
\label{sec:setup}
\subsection{AI-code Generation}
To perform code generation and assess the tool on the AI-generated code, we adopted four state-of-the-art NMT models.

\noindent
$\blacksquare$ \textbf{Seq2Seq} is a model that maps an input of sequence to an output of sequence. 
Similar to the encoder-decoder architecture with attention mechanism \cite{bahdanau2014neural}, we use a bi-directional LSTM as the encoder to transform an embedded intent sequence into a vector of hidden states with equal length. 
We implement the Seq2Seq model using \textit{xnmt}~\cite{neubig18xnmt}. 
We use an Adam optimizer \cite{kingma2015adam} with $\beta_1=0.9$ and $\beta_2=0.999$, while the learning rate $\alpha$ is set to $0.001$. We set all the remaining hyper-parameters in a basic configuration: layer dimension = $512$, layers = $1$, epochs = $200$, beam size = $5$.

\noindent
$\blacksquare$ \textbf{CodeBERT}~\cite{feng2020codebert} is a large multi-layer bidirectional Transformer architecture~\cite{vaswani2017attention} pre-trained on millions of lines of code across six different programming languages. 
Our implementation uses an encoder-decoder framework where the encoder is initialized to the pre-trained CodeBERT weights, and the decoder is a transformer decoder, composed of $ 6$ stacked layers. The encoder follows the RoBERTa architecture~\cite{DBLP:journals/corr/abs-1907-11692}, with $12$ attention heads,  hidden layer dimension of $768$, $12$ encoder layers, and $514$ for the size of position embeddings. We set the learning rate $\alpha = 0.00005$, batch size = $32$, and beam size = $10$.

\noindent
$\blacksquare$ \textbf{CodeT5+}~\cite{wang2023codet5+} is a new family of Transformer models pre-trained with a diverse set of pretraining tasks including causal language modeling, contrastive learning, and text-code matching to learn rich representations from both unimodal code data and bimodal code-text data. 
We utilize the variant with model size $220M$, which is trained from scratch following T5’s architecture~\cite{DBLP:journals/jmlr/RaffelSRLNMZLL20}. It has an encoder-decoder architecture with $12$ decoder layers, each with $12$ attention heads and hidden layer dimension of $768$, and $512$ for the size of position embeddings. We set the learning rate $\alpha = 0.00005$, batch size = $16$, and beam size = $10$.

\noindent
$\blacksquare$ \textbf{PLBart}~\cite{PLBart} is a multilingual encoder-decoder (sequence-to-sequence) model primarily intended for code-to-text, text-to-code, code-to-code tasks. The model is pre-trained on a large collection of Java and Python functions and natural language descriptions collected from GitHub and StackOverflow.
We use the PLBart-large architecture with $12$ encoder layers and $12$ decoder layers, each with $16$ attention heads. We set the learning rate $\alpha = 0.00005$, batch size = $16$, and beam size = $10$.

\begin{figure}
    \centering
    \includegraphics[width=\linewidth]{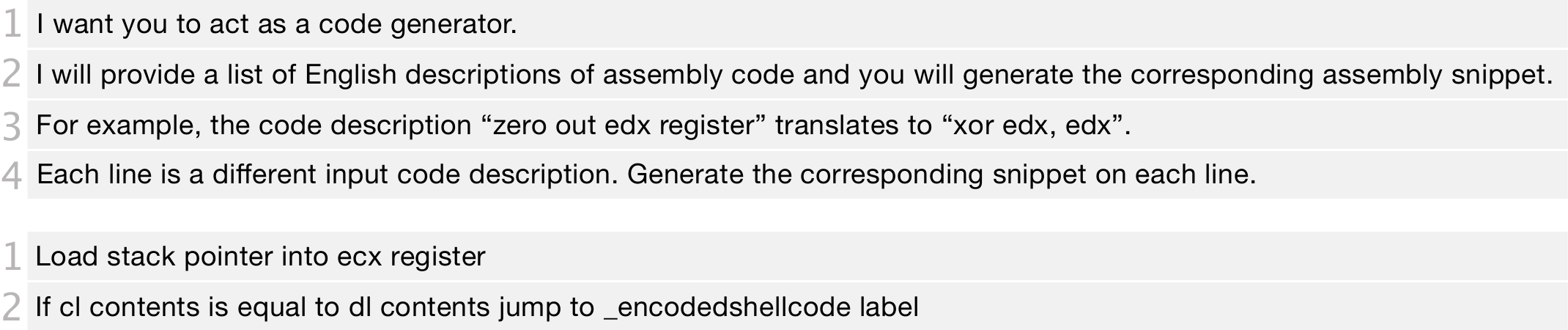}
    \caption{Example of prompt used for code generation with ChatGPT. 
    }
    \label{fig:gen_prompt_example}
\end{figure}

\noindent
$\blacksquare$ \textbf{ChatGPT}~\cite{chatgpt} is a popular large language model by OpenAI which extends the GPT model series. It can provide detailed responses given natural language prompts and examples of the desired output. 
We used the latest free version of the ChatGPT-3.5 model~\cite{chatgptversion}, which we queried via web interface and was trained on data up to April 2023. 
We prompted ChatGPT using the query format illustrated in~\figurename{}~\ref{fig:gen_prompt_example}. We first provided a detailed description of the required task, i.e., the generation of assembly code snippets starting from NL descriptions, and included an example of input and the desired output. Then, we provided a list of natural language code descriptions and asked ChatGPT to automatically generate the corresponding assembly instructions. 
We set the \textit{temperature} parameter, which controls the degree of model randomness, at its default value of $1$. By adopting the default configuration, we test the model’s typical behavior, without forcing it to be more predictable or more creative~\cite{buscemi2023comparative}.

To choose the models' hyperparameters that differ from the standard configuration~\cite{yang2023exploitgen, liguori2021evil} we followed the state-of-the-art in automatic exploit generation (see~\S{}~\ref{sec:related}). We also evaluated the validation set every 320 steps.

We followed the best practices in the field of code generation by supporting NMT models with data processing operations.
The data processing steps are usually performed both before translation (\textit{pre-processing}), to train the NMT model and prepare the input data, and after translation (\textit{post-processing}), to improve the quality and the readability of the code in output.

Our pre-processing operations start with the \textit{stopwords filtering}, i.e., we remove a set of custom-compiled words (e.g., \textit{the}, \textit{each}, \textit{onto}) from the intents to include only relevant data for machine translation. 
Next, we use a \textit{tokenizer} to break the intents into chunks of text containing space-separated words (i.e., the \textit{tokens}). 
To improve the performance of the machine translation~\cite{li2018named,modrzejewski2020incorporating,liguori2021evil}, we \textit{standardize} the intents (i.e., we reduce the randomness of the NL descriptions) by using a \textit{named entity tagger}, which returns a dictionary of \textit{standardizable} tokens, such as specific values, label names, and parameters, extracted through regular expressions. We replace the selected tokens in every intent with ``\textit{var}\#", where \# denotes a number from $0$ to \textit{$|l|$}, and $|l|$ is the number of tokens to standardize. 
Finally, the tokens are represented as real-valued vectors using \textit{word embeddings}. 

The pre-processed data is used to feed the NMT model. Once the model is trained, we perform the code generation from the NL intents. Therefore, when the model takes as inputs new intents, it generates the related code snippets based on its knowledge (\textit{model's prediction}).
As for the intents, also the code snippets predicted by the models are processed (\textit{post-processing}) to improve the quality and readability of the code. Finally, the dictionary of standardizable tokens is used in the \textit{de-standardization} process to replace all the ``\textit{var}\#" with the corresponding values, names, and parameters.

In the data pre-processing phase, we employ the \textit{nltk word tokenizer}~\cite{bird2006nltk} to tokenize the NL intents and the Python \textit{tokenize} package~\cite{tokenize} for the code snippets. 
To facilitate the standardization of NL intents, we implement a named entity tagger using \emph{spaCy}, an open-source, NL processing library written in Python and Cython~\cite{spacy}.

\subsection{Dataset}

\begin{table}[t]
\centering
\caption{Dataset statistics}
\label{tab:dataset_statistics}
\footnotesize
\begin{tabular}{
>{\centering\arraybackslash}m{4cm} |
>{\centering\arraybackslash}m{4cm}
>{\centering\arraybackslash}m{4cm}}
\toprule
\textbf{Metric} & \textbf{NL Intents} & \textbf{Assembly Code Snippets}\\ \toprule
\textit{Unique lines}    & $5,740$   & $3,316$\\
\textit{Unique tokens}      & $2,855$   & $1,770$\\
\textit{Avg. tokens per line}        & $9.18$    & $5.83$\\
\bottomrule
\end{tabular}
\end{table}

To feed the models for the generation of security-oriented code, we extended the publicly available \textit{Shellcode\_IA32} dataset~\cite{liguori2021shellcode_ia32,liguori2022can} for automatically generating \textit{shellcodes} from NL descriptions. 
A shellcode is a list of machine code instructions to be loaded in a vulnerable application at runtime. 
The traditional way to develop shellcodes is to write them using the assembly language, and by using an assembler to turn them into \emph{opcodes} (operation codes, i.e., a machine language instruction in binary format, to be decoded and executed by the CPU) \cite{foster2005sockets,megahed2018penetration}. 
Common objectives of shellcodes include spawning a system shell, killing or restarting other processes, causing a denial-of-service (e.g., a fork bomb), leaking secret data, etc.

The dataset consists of instructions in assembly language for \textit{IA-32} collected from publicly available security exploits~\cite{exploitdb,shellstorm}, manually annotated with detailed English descriptions. In total, it contains $3,200$ unique pairs of assembly code snippets/English intents. 
We further enriched the dataset with additional samples of shellcodes collected from publicly available security exploits, reaching $5,900$ unique pairs of assembly code snippets/English intents. To the best of our knowledge, the resulting dataset is the largest collection of shellcodes in assembly available to date.

Our dataset also includes $1,374$ intents (${\sim}23\%$ of the dataset) that generate multiple lines of assembly code, separated by the newline character \textit{\textbackslash{n}}. These multi-line snippets contain many different assembly instructions (e.g., whole functions) that can range between $2$ and $18$.
For example, the copy of the ASCII string \textit{``/bin//sh"} into a register is a typical operation to spawn a shell, which requires three distinct assembly instructions: push the hexadecimal values of the words \textit{``/bin"} and \textit{``//sh"} onto the stack register before moving the contents of the stack register into the destination register. 
Further examples of multi-line snippets include conditional jumps, tricks to zero out the registers without generating null bytes, etc. 
Table~\ref{tab:multi_line} shows two further examples of multi-line snippets with their natural language intents. 

\begin{table}[t]
\caption{Examples of assembly code with NL descriptions from our dataset.}
\label{tab:multi_line}
\footnotesize
\begin{tabular}
{>{\centering\arraybackslash}m{6.25cm} | 
>{\centering\arraybackslash}m{6.25cm}}
\toprule
\textbf{Code Snippet} & \textbf{English Intent} \\ \midrule
\texttt{xor bl, 0xBB} \textbackslash{n} \texttt{jz formatting} \textbackslash{n}
\texttt{mov cl, byte [esi]}
& \textit{Perform the xor between BL register and 0xBB and jump to the label formatting if the result is zero else move the current byte of the shellcode in the CL register.}\\ \midrule
\texttt{xor ecx, ecx} \textbackslash{n} \texttt{mul ecx} &
\textit{Zero out the EAX and ECX registers.
}\\ \bottomrule
\end{tabular}
\end{table}

Table~\ref{tab:dataset_statistics} summarizes the statistics of the dataset used in this work, including the unique examples of NL intents and assembly code snippets, the unique number of tokens, and the average number of tokens per snippet and intent.
The dataset is publicly available on GitHub\footnote{\url{https://github.com/dessertlab/Shellcode_IA32}}.

To perform the experiments, we split the dataset into training, validation, and test sets using a common $80\%/10\%/10\%$ ratio~\cite{kim2018artificial,DBLP:conf/msr/MashhadiH21}. Hence, they contain $4720$, $590$ and $590$ samples, respectively.

\subsection{Baseline Assessment Solutions}
\label{subsec:metrics}

As a baseline for the evaluation, we used the following output similarity metrics, which are widely used to assess the performance of AI generators in many code generation tasks~\cite{LIGUORI2023120073}, including the generation of assembly code for security contexts~\cite{yang2022dualsc,yang2023exploitgen,ruan2023prompt,liguori2021evil,liguori2022can}:

\begin{itemize}

    \item \textbf{Compilation Accuracy (CA)}. It indicates whether each code snippet produced by the model is compilable according to the syntax rules of the target language. CA value is either $1$, when the snippet's syntax is correct, or $0$ otherwise. To compute the \textit{compilation accuracy}, we used the \textit{Netwide Assembler} (NASM) assembler~\citep{nasm}.

    \item \textbf{Bilingual Evaluation Understudy (BLEU) score}~\cite{papineni2002bleu}. It measures the degree of n-gram overlapping between the string of each code snippet produced by the model and the reference. This metric also takes into account a \textit{brevity penalty} to penalize predictions shorter than the references. 
    BLEU value ranges between $0$ and $1$, with higher scores corresponding to a better quality of the prediction.  Similar to previous studies, we use the BLEU-4 score (i.e., we set $n=4$).
    We implemented BLEU score computation employing the \texttt{bleu\_score} module contained in the open-source Python suite Natural Language Toolkit (NLTK)~\cite{bleu}. 

    \item \textbf{SacreBLEU}~\cite{post-2018-call}. This is a different implementation of the BLEU score which differs from the traditional one because it uses different tokenization techniques. We used the implementation available on Hugging Face~\cite{SacreBLEU}



    \item \textbf{Exact Match accuracy (EM)}. It indicates whether each code snippet produced by the model perfectly matches the reference. EM value is $1$ when there is an exact match, $0$ otherwise.
    To compute the exact match, we used a simple Python string comparison.

    \item \textbf{Edit Distance (ED)}. It measures the \textit{edit distance} between two strings, i.e., the minimum number of operations on single characters required to make each code snippet produced by the model equal to the reference. ED value ranges between $0$ and $1$, with higher scores corresponding to smaller distances. 
    For the edit distance, we adopted the Python library \texttt{pylcs}~\cite{pylcs}. 

\end{itemize}

\begin{figure}[ht]
    \centering
    \includegraphics[width=0.8\linewidth]{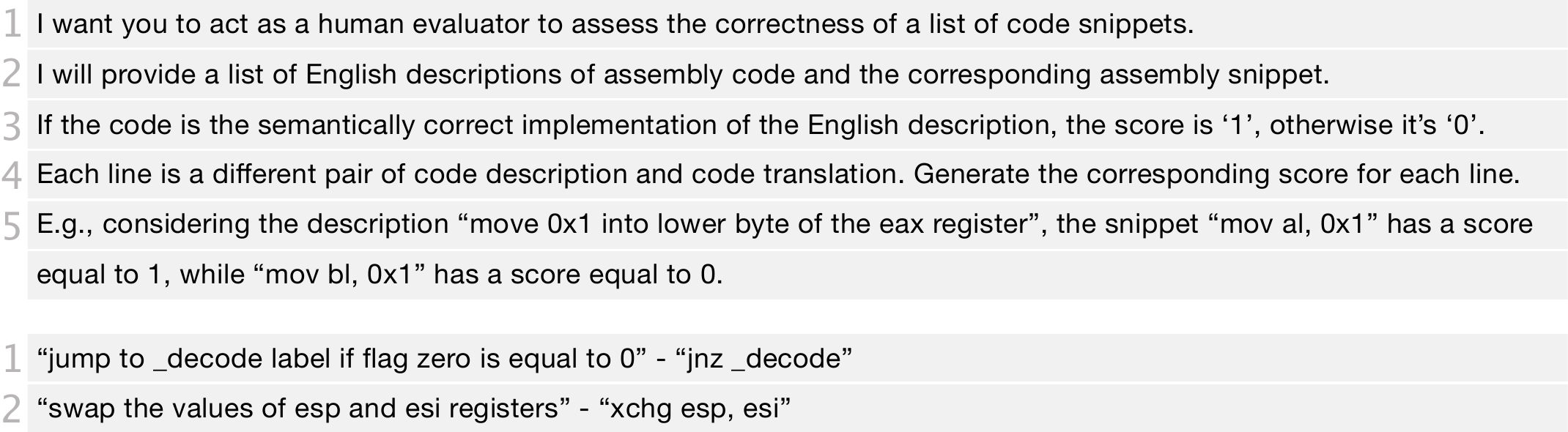}
    \caption{Example of prompt used for code correctness assessment with respect to the English description using ChatGPT.\\}
    \label{fig:GPT-NL}
    \includegraphics[width=0.8\linewidth]{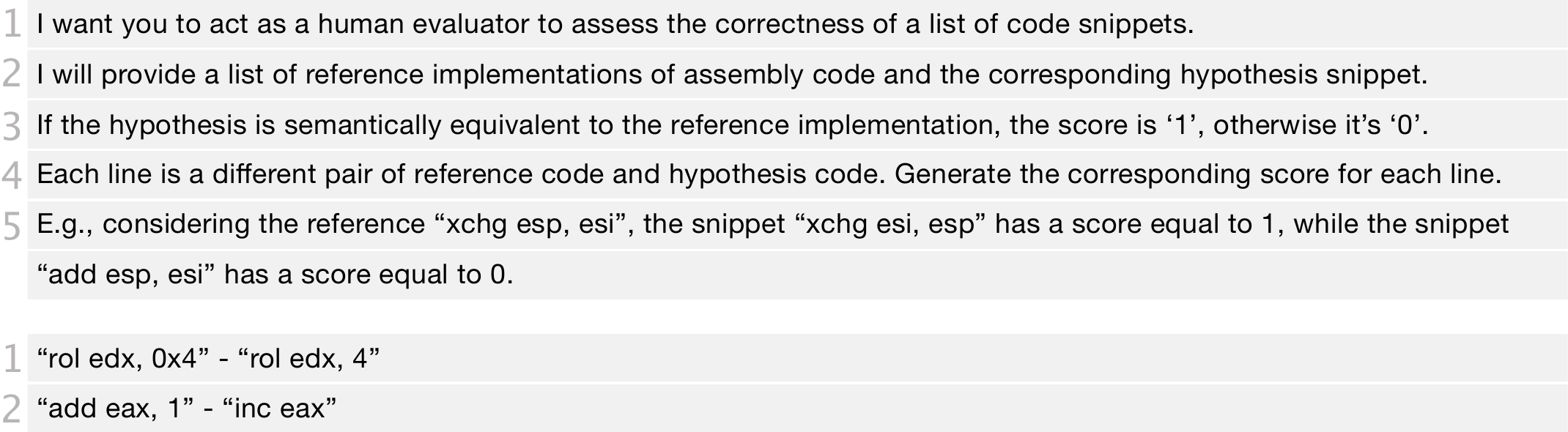}
    \caption{Example of prompt used for code correctness assessment with respect to the reference implementation using ChatGPT.}
    \label{fig:GPT-GT}
\end{figure}

As a further baseline for the comparison, we again adopted the famous \textbf{ChatGPT}~\cite{chatgpt}. As for the code generation process, we used the latest free version of the ChatGPT-3.5 model~\cite{chatgptversion}, which we queried via the web interface and was trained on data up to April 2023. We kept the default temperature value equal to 1 to test the model’s typical behavior.

For every code generated by the models, we asked ChatGPT to assess the code by assigning a value of $1$ when it is correct, $0$ otherwise. 
We performed two different evaluations:

\begin{itemize}
    \item \textbf{ChatGPT-NL}: ChatGPT evaluates if the code generated by the models is the translation in assembly code of what is required in the natural language intents, similar to what a human evaluator does during the manual code review;
    \item \textbf{ChatGPT-GT}: ChatGPT evaluates if the code generated by the models is semantically equivalent to the ground truth used as a reference for the evaluation, similar to the assessment performed by output similarity metrics.
\end{itemize}

The prompt used to query the model in both settings are shown in~\figurename{}~\ref{fig:GPT-NL} and~\ref{fig:GPT-GT}, respectively. For the ChatGPT-NL evaluation, we provided a list of NL code descriptions and the corresponding code snippet generated by each model and asked ChatGPT to assign a score of $1$ when the code is the semantically correct implementation of the description, $0$ otherwise. For the ChatGPT-GT evaluation, we provided a list of reference implementations of the code snippets, i.e., the ground truth, and the corresponding code snippet generated by each model, and asked ChatGPT to assign a score of $1$ when the generated code is semantically equivalent to the reference code, $0$ otherwise.

Different from output similarity metrics and our tool, ChatGPT does not have a deterministic behavior. Therefore, we executed multiple assessments of code correctness with the OpenAI model by performing three repetitions. We found that in the ~97\% of the cases, there were no discrepancies among the different repetitions. In the few discrepancy cases, since the correctness score is 1 (correct) or 0 (incorrect), we adopted a majority voting mechanism (i.e., 2 out of 3).

\subsection{Human Evaluation}
\label{subsec:ground_truth}



To ensure a robust and thorough assessment of both \approach{} and baseline approaches in evaluating AI-generated code, we conducted a meticulous comparison against human evaluation, which serves as ground truth for our analysis.

In the human evaluation, a code prediction was deemed successful if it accurately translated the NL description into the assembly language and adhered to the established assembly programming language rules, warranting a score of $1$. Any deviation resulted in a score of $0$.

The manual evaluation was conducted by a diverse group consisting of $3$ human evaluators, all with a computer science background and expertise in assembly language and cybersecurity. The group included individuals with varying degrees of professional experience and educational qualifications. In particular, $2$ Ph.D. students with a master's degree and a researcher with a Ph.D. in information technologies. The diversity and expertise of our evaluators ensured the reliability of our human evaluation process.

To fortify the integrity of our evaluation and minimize the potential for human error, the 3 human evaluators independently scrutinized each code snippet generated by the models. 
Any discrepancies that arose were attributed to human oversight and promptly rectified, culminating in unanimous consensus across all cases in the human evaluation, demonstrating a resounding $100\%$ alignment.

\section{Experimental Results}
\label{sec:results}
For our experiments, we used a machine with a Debian-based distribution, with 8 vCPU, 16 GB RAM, and one Nvidia T4 GPU.

In the following, we assess how \approach{} aligns with the human evaluation for the correctness of the code generated by models and compare the results with the baseline assessment solutions by performing both a quantitative (\S{}~\ref{subsec:quantitative}) and correlation analysis (\S{}~\ref{subsec:correlation}).
We also show representative examples to investigate discrepancy cases between \approach{} and the human evaluation (\S{}~\ref{subsec:qualitative}). 
Finally, we assess the computational cost of running \approach{} on the code predictions to provide an estimate of the average time needed to assess code correctness (\S{}~\ref{subsec:cost}).

\subsection{Quantitative Analysis}
\label{subsec:quantitative}
For all the models, we compared the average code correctness values computed by \approach{} over all the examples of the test set with respect to the average semantic correctness assessed with the human evaluation and the alignment between the two solutions.
Table~\ref{tab:quantitative} shows the results.

The quantitative analysis is of limited use if we do not consider the matching rate between \approach{} and the human evaluation, i.e., the percentage of code snippets that are equivalently considered correct or incorrect by both methods.
For instance, both \approach{} and the human evaluation can assess the code snippets as correct for 50\% of the cases but still have 100\% of discrepancy cases (i.e., the code snippets considered correct by \approach{} and human evaluation are disjoint sets). 
The table highlights that the results provided by \approach{} are very close to the human evaluation. Indeed, our method and human evaluation provide, on average, the same classification in the $\sim93\%$ of models' predictions (min $90\%$, max $95\%$). 
These results suggest that the proposed approach aligns well with human evaluations.

The table also shows the results of \approach{} and the human evaluation for each model. The method classifies, on average, $65\%$ of the generated code snippets by all the models as correct. On the other hand, according to our manual code review, we found that, on average, the $71\%$ of the generated code snippets are semantically correct. Hence, although they are very close, \approach{} underestimates the code correctness when compared to human evaluation. This is because human evaluators are not always able to account for all the factors that contribute to the program state. For example, the instructions \texttt{ADD EAX, 1} and \texttt{INC EAX} perform a similar operation, i.e., incrementing the \texttt{EAX} register, yet alter different flags, i.e., \texttt{INC} does not affect the Carry flag, hence they are not equivalent.
Moreover, some snippets exceeded the constraint we imposed on the maximum number of symbolic execution steps to perform (i.e., \textit{max\_step}), hence leading to a correctness score equal to $0$. However, it is important to notice that these cases represent a minimal fraction of the total examples ($16$ out of $2950$ predictions, i.e., the 0.54\%).

\begin{table}[t]
\centering
\caption{Code correctness assessment of \approach{} with respect to the human evaluation.}
\label{tab:quantitative}
\footnotesize
\begin{tabular}{
>{\centering\arraybackslash}m{2.5cm} |
>{\arraybackslash}m{1.25cm}
>{\arraybackslash}m{1.75cm}
>{\arraybackslash}m{1.5cm} 
>{\arraybackslash}m{1.25cm} 
>{\arraybackslash}m{1.65cm} |
>{\arraybackslash}m{1.25cm}
}
\toprule
& \textbf{Seq2Seq} & \textbf{CodeBERT} & \textbf{CodeT5+} & \textbf{PLBart} & \textbf{ChatGPT} & \textbf{Average} \\ \toprule
\textit{Matching Rate} & \Chart{0.94} & \Chart{0.90} & \Chart{0.93} & \Chart{0.92} & {\Chart{0.95}} & {\Chart{0.93}}\\
\approach{} & \Chart{0.60} & \Chart{0.68} & \Chart{0.69} & \Chart{0.57} & {\Chart{0.71}} & {\Chart{0.65}}\\
\textit{Human Eval.} & \Chart{0.66} & \Chart{0.78} & \Chart{0.76} & \Chart{0.64} & \Chart{0.73} & {\Chart{0.71}}\\    

\bottomrule
\end{tabular}
\end{table}

To better appreciate the evaluation provided by \approach{}, we compared the results of the human evaluation with the results provided by the baseline solutions (described in \S{}~\ref{subsec:metrics}). 
To this aim, we computed an \textit{offset value}, i.e., the difference between the optimal value represented by the human evaluation and the value provided by different assessing solutions. The lower the offset, the closer the result is to the human evaluation. Table~\ref{tab:offset} shows the results.

\begin{table}[t]
\centering
\caption{Offset with respect to the human evaluation. The best performance (lower offset) is \textcolor{blue}{blue}, while the worst performance (higher offset) is \textcolor{red}{red}.}
\label{tab:offset}
\footnotesize
\begin{tabular}{ 
>{\arraybackslash}m{2.5cm} |
>{\arraybackslash}m{1.25cm}
>{\arraybackslash}m{1.75cm}
>{\arraybackslash}m{1.5cm}
>{\arraybackslash}m{1.25cm} 
>{\arraybackslash}m{1.75cm} |
>{\arraybackslash}m{1.25cm}}
\toprule
\textbf{Evaluation} & \textbf{Seq2Seq} & \textbf{CodeBERT} & \textbf{CodeT5+} & \textbf{PLBart} & \textbf{ChatGPT} & \textbf{Average}\\ \midrule

\textit{Compilat. Acc.} & \textcolor{red}{\Chart{0.28}} & \Chart{0.15} & \Chart{0.14} & \Chart{0.21} & \Chart{0.24} & \Chart{0.20}\\
\textit{BLEU-4} & \Chart{0.22} & \textcolor{red}{\Chart{0.20}} & \textcolor{red}{\Chart{0.29}} & \textcolor{red}{\Chart{0.24}} & \textcolor{red}{\Chart{0.29}} & \textcolor{red}{\Chart{0.25}}\\
\textit{SacreBLEU} & \Chart{0.15} & \Chart{0.05} & \textcolor{blue}{\Chart{0.07}} & \Chart{0.09} & \Chart{0.09} & \Chart{0.09}\\
\textit{Edit Distance} & \Chart{0.15} & \Chart{0.07} & \textcolor{blue}{\Chart{0.07}} & \Chart{0.09} & \Chart{0.05} & \Chart{0.08}\\
\textit{Exact Match} & \Chart{0.25} & \Chart{0.17} & \Chart{0.27} & \textcolor{red}{\Chart{0.24}} & \Chart{0.28} & \Chart{0.24}\\
\textit{ChatGPT-NL} & \Chart{0.08} & \textcolor{blue}{\Chart{0.02}} & \textcolor{blue}{\Chart{0.07}} & \Chart{0.17} & \Chart{0.24} & \Chart{0.12}\\
\textit{ChatGPT-GT} & \textcolor{blue}{\Chart{0.06}} & \Chart{0.13} & \Chart{0.10} & \Chart{0.10} & \Chart{0.05} & \Chart{0.09}\\
\approach{} & \textcolor{blue}{\Chart{0.06}} & \Chart{0.10} & \textcolor{blue}{\Chart{0.07}} & \textcolor{blue}{\Chart{0.07}} & \textcolor{blue}{\Chart{0.02}} & \textcolor{blue}{\Chart{0.06}}\\
\bottomrule
\end{tabular}
\end{table}

The average offset of the output similarity metrics ranges between a minimum (best) value of $0.08$ (for edit distance) and a maximum (worst) value of $0.25$ (for BLEU-4).
ChatGPT provided results similar to the best-performing output similarity metrics, with an average offset equal to $0.09$ over all the models when the correctness of models' predictions is computed with respect to the NL intent (ChatGPT-NL), and equal to $0.12$ when the predictions are compared to the ground truth (ChatGPT-GT).
\approach{} provided the lowest offset in $4$ out of $5$ models and, an average offset equal to $0.06$, which results to be the lowest value, i.e., the code correctness computed by the proposed method is, on average, the closest to the human evaluation.

In our experiments and analysis, we observed that compilation errors encountered during the assembling process were exclusively attributable to the syntactic inaccuracies within the code snippets generated by the models. The occurrence of compilation errors due to syntactical inaccuracies underscores the challenges of AI code generation models in generating assembly code.

\subsection{Qualitative Analysis}
\label{subsec:qualitative}
We performed a manual inspection of the cases of discrepancy to examine when the method provides different results from the human evaluation.
We have a discrepancy case when the method assesses the code as correct but the human evaluation does not, or when the method assesses the code as incorrect it is semantically correct according to the human evaluation.

As shown in \tablename{}~\ref{tab:quantitative}, the method underestimates the performance of the models. In fact, an in-depth inspection of the results revealed that $\sim99\%$ of the discrepancy cases were due to examples classified as correct by human evaluation (value $1$) but incorrect \approach{} (value $0$). 
To better discuss these discrepancy cases, Table~\ref{tab:qualitative_sem} illustrates four representative examples of mismatch between \approach{} and the human evaluation.

\begin{table}[t]
\centering
\caption{Examples of mismatches \approach{} and human evaluation.}
\label{tab:qualitative_sem}
\footnotesize
\begin{tabular}{
>{\arraybackslash}m{3.5cm} |
>{\centering\arraybackslash}m{2.5cm}
>{\centering\arraybackslash}m{2.5cm}
>{\centering\arraybackslash}m{1.5cm}
>{\centering\arraybackslash}m{1.5cm}}
\toprule
\textbf{NL Intent} & \textbf{Ground Truth} & \textbf{Model's Prediction} & \textbf{Human Eval.} & \textbf{\approach{}}  \\ 
\midrule
\textit{Push zero into the stack} & \texttt{xor edx, edx} \textbackslash n \texttt{push edx} & \texttt{xor eax, eax} \textbackslash n \texttt{push eax} & 1 & 0\\
\midrule
\textit{Save eax contents in esi contents} & \texttt{mov esi, eax} & \texttt{xchg esi, eax} & 1 & 0\\
\midrule 
\textit{In L1 jump short to esp}  & \texttt{L1: jmp short esp}  & \texttt{jmp short esp} & 0 & 1 \\
\midrule 
\textit{Restore the top of the stack into the ecx register then decrement the ecx register and jump to the l1 label if the contents of the ecx register is not zero} & \texttt{pop ecx} \textbackslash n \texttt{loop l1} & \texttt{pop ecx} \textbackslash n \texttt{dec ecx} \textbackslash n \texttt{jmp l1} & 0 & 1\\
\bottomrule
\end{tabular}
\end{table}

The first two rows of the table showcase two model predictions that are correctly labeled by human evaluation, but considered incorrect by our method. These misclassifications were due to the ambiguity of the code snippets since the same NL description can be expressed by semantically different code snippets. For instance, to zero out the stack (row \# 1), a programmer can reset any register and then push the contents of the register (i.e., $0$) into the stack register. Also, to move the contents of a register into a different one (row \#2), a programmer can use the \texttt{mov} instruction to transfer a value from a source to a destination, or, equivalently, the \texttt{xchg} instruction, to swap the contents of the registers. Both the code snippets generated by the model accomplish what is required in the NL intent, but at the end of the symbolic execution, the state of the registers is different from the one obtained with the code in the ground truth (\texttt{EAX} is reset instead of \texttt{EDX} in the row \# 1, while, in row \# 2, \texttt{EAX} contains the value of \texttt{ESI}, instead of his original value). Therefore, \approach{} provides the SEM score equal to zero, even if the snippets are semantically correct.

The last two rows of the table show examples of incorrect predictions that are wrongly classified as correct by the tool.
As already remarked, these cases are very limited in numbers and represent situations in which, although the symbolic execution of predictions and ground-truth reference lead to the same state of the registers at the end of the execution, the model's prediction is not what is required by the NL description. 
For instance, in row \# 3, the prediction contains what is described in the NL intents except for the label \texttt{L1}. The label does not affect the state of the registers during the execution of the code, but it will impact the behavior of the whole program (unless the label is never used by other instructions).  
Row \#4, instead, showcases a more complex example in which the correct instruction \texttt{loop}, which decrements the value of the counter register \texttt{ECX} and jumps to the args (i.e., the \texttt{l1} label) if the counter is not zero, is replaced, in the model's prediction, by the decrement of the counter (\texttt{dec ecx} and an unconditional jump (\texttt{jmp}). In this case, although the instructions led to the same state of the registers because the counter was not zero after the decrement, the prediction is incorrect since the unconditional jump does not take into account the condition on the \texttt{ecx} register specified in the NL intent.


\subsection{Correlation Analysis}
\label{subsec:correlation}
Additionally, we performed a statistical analysis by computing the correlation of \approach{} with the human evaluation overall the code snippets of the test set (i.e., we considered the values on the single predictions). 

To this aim, we computed the \textit{Pearson} correlation coefficient $r$, which measures the strength of association (i.e., the linear relationship) between two variables in a correlation analysis and is defined as the covariance of the two variables divided by the product of their respective standard deviations~\cite{pearson1895notes}. 
The correlation coefficient is a unit-free value between $-1$ and $1$, which represents \textit{perfect} correlation, \textit{negative}, and \textit{positive}, respectively.
Positive values indicate a positive correlation, i.e., the values of both variables tend to increase together, while negative values indicate a negative correlation, i.e., the values of one variable tend to increase when the values of the other variable decrease.
A high value of the coefficient indicates that there is a strong correlation with the human evaluation. On the contrary, a small value indicates that there is a weak correlation with human evaluation.
To provide context for the evaluation, we also computed the correlation coefficients between the baseline solutions and the human evaluation. \tablename{}~\ref{tab:correlation} shows the results.

\begin{table}[t]
\centering
\caption{Pearson correlation coefficient $r$ with human evaluation. For every model, the best values are \textcolor{blue}{blue}, while the worst are \textcolor{red}{red}.}
\label{tab:correlation}
\footnotesize
\begin{tabular}{ 
>{\arraybackslash}m{2.5cm} |
>{\arraybackslash}m{1.25cm}
>{\arraybackslash}m{1.75cm}
>{\arraybackslash}m{1.5cm}
>{\arraybackslash}m{1.25cm} 
>{\arraybackslash}m{1.75cm}|
>{\arraybackslash}m{1.25cm}}
\toprule
\textbf{Evaluation} & \textbf{Seq2Seq} & \textbf{CodeBERT} & \textbf{CodeT5+} & \textbf{PLBart} & \textbf{ChatGPT} & \textbf{Average}\\ \midrule
\textit{Compilat. Acc.} & \textcolor{red}{\Chart{0.36}} & \Chart{0.44} & \Chart{0.59} & \Chart{0.52} & {\Chart{0.22}} & {\Chart{0.43}}\\
\textit{BLEU-4} & \Chart{0.56} & \Chart{0.55} & \Chart{0.56} & \Chart{0.64} & {\Chart{0.50}} & {\Chart{0.56}}\\
\textit{SacreBLEU} & \Chart{0.53} & \Chart{0.52} & \Chart{0.60} & \Chart{0.66} & {\Chart{0.51}} & {\Chart{0.57}}\\
\textit{Edit Distance} & \Chart{0.67} & \Chart{0.61} & \Chart{0.75} & \Chart{0.76} & {\Chart{0.64}} & {\Chart{0.69}}\\
\textit{Exact Match} & \Chart{0.60} & \Chart{0.67} & \Chart{0.55} & \Chart{0.61} & {\Chart{0.55}} & {\Chart{0.59}}\\
\textit{ChatGPT-NL} & \Chart{0.42} & \textcolor{red}{\Chart{0.42}} & \textcolor{red}{\Chart{0.37}} & \textcolor{red}{\Chart{0.44}} & \textcolor{red}{{\Chart{0.07}}} & \textcolor{red}{{\Chart{0.34}}}\\
\textit{ChatGPT-GT} & \Chart{0.67} & \Chart{0.61} & \Chart{0.70} & \Chart{0.68} & {\Chart{0.56}} & {\Chart{0.64}}\\
\approach{} & \textcolor{blue}{\Chart{0.87}} & \textcolor{blue}{\Chart{0.78}} & \textcolor{blue}{\Chart{0.85}} & \textcolor{blue}{\Chart{0.85}} & \textcolor{blue}{{\Chart{0.87}}} & \textcolor{blue}{{\Chart{0.84}}}\\
\bottomrule
\end{tabular}
\end{table}

Confirming previous work~\cite{LIGUORI2023120073}, the analysis shows that Edit Distance and Exact Match are the output similarity metrics most correlated to the semantic correctness for security-oriented code, with both Pearson's $r$ coefficients equal to $0.69$ and $0.59$, respectively. The output similarity metric that is less correlated to human evaluation is the compilation accuracy ($r=0.43$), showing that the syntactic correctness of the code is not highly correlated to its semantic correctness.

An important takeaway from our experiments is that, despite ChatGPT-based assessments providing results close to the human evaluation in the quantitative analysis (see \ref{tab:offset}), they have a correlation coefficient lower than the best-performing output similarity metric. Indeed, ChatGPT-GT has a correlation coefficient equal to $0.64$, while ChatGPT-NL has a very poor correlation with human evaluation, resulting in the lowest value among all the baseline solutions ($r=0.34$). This is a consequence of the high number of discrepancy cases between these solutions and the human evaluation, which is even more exacerbated in the ChatGPT-NL solution.

Finally, \approach{} provides the highest correlation coefficient over all the five models, with an average value $r=0.84$, hence being the only one to have a \textit{very strong} correlation with human evaluation~\cite{akoglu2018user}, surpassing by far all other metrics for all models.

\subsection{Computational Cost}
\label{subsec:cost}
We assessed the computational cost of \approach{} in assessing the code correctness.
Since the method skips the symbolic execution process for all generated snippets that are identical to the ground truth or that are not syntactically correct, we performed a thorough analysis considering three different cases: the evaluation of all the predictions in the test set (i.e., $590$ code snippets), the evaluation of the subset of generated snippets that do not match the ground truth (i.e., ``PR $\neq$ GT''), and evaluation of the subset of generated snippets that do not match the ground truth and are also syntactically correct (i.e., ``PR $\neq$ GT \& SYN=1'').

\tablename~\ref{tab:cost_per_snippet} presents a detailed analysis of the computational cost of \approach{}. The table shows the average, median, and standard deviation of our method's cost, in terms of time (seconds) to assess a single code snippet, for each model.
The last row shows the average, standard deviation, and median of our method's cost per snippet, across all models.

Regarding the evaluation of the entire test set, the average time required to assess a snippet is, as expected, lower than in other cases, with an average time equal to $\sim0.20$s, a standard deviation of $\sim1.28$s and a median value of $0.0$. This is because predictions matching the ground truth are included in the analysis. In fact, in these cases, both syntactic assessment and symbolic execution for the semantic assessment are skipped.

When we consider the subset of samples in which PR $\neq$ GT, i.e. when we exclude from our analysis the predictions that perfectly match the ground truth, the mean time per snippet increases on all five models. Indeed, in this case, \approach{} requires on average $\sim0.37$s to assess the code correctness, with a standard deviation of $\sim1.73$s and a median value of $\sim0.21$s.

In the last scenario, PR $\neq$ GT \& SYN=1, the value increases again because snippets that were labeled as syntactically incorrect during the assembling process (see \S{}~\ref{assembling_process}) are excluded from the analysis. 
Therefore, this evaluation concerns only the predictions that went through the symbolic execution, i.e., all the evaluation steps of the proposed method.
Interestingly, the mean time to assess the correctness of a code snippet produced by ChatGPT is slightly higher than other models. This is because the other models generate less syntactically correct code snippets than ChatGPT, which are therefore excluded from the analysis, i.e., $553$, $550$, $531$, $499$ for Seq2Seq, CodeBERT, CodeT5+ and PLBart, respectively, against the $577$ by ChatGPT (out of $590$, i.e., the examples in the test set).
Overall, regardless of the model, \approach{} needs on average $0.43$s to symbolically execute the code and perform the evaluation, with a standard deviation of $\sim1.87$s and a median value of $\sim0.22$s.

\begin{table}[t!]
\caption{Avg., median and std. dev. of the computational time (seconds) of \approach{} per snippet on the whole data, the subset of predictions not matching the reference, and the subset that is also syntactically correct.}
\label{tab:cost_per_snippet}
\centering
\footnotesize
\begin{tabular}{
>{\arraybackslash}m{1.8cm} |
>{\centering\arraybackslash}m{1.75cm}
>{\centering\arraybackslash}m{1.5cm} |
>{\centering\arraybackslash}m{1.75cm}
>{\centering\arraybackslash}m{1.5cm} |
>{\centering\arraybackslash}m{1.75cm}
>{\centering\arraybackslash}m{1.5cm} }  
\toprule & 
\multicolumn{2}{c|}{\textbf{All predictions}} & \multicolumn{2}{c|}{\textbf{PR$\neq$ GT}} & \multicolumn{2}{c}{\textbf{PR $\neq$ GT $\&$ SYN=1}} \\ 
&\textit{ Avg \& SD} & \textcolor{blue}{\textit{Median}} & \textit{Avg \& SD} & \textcolor{blue}{\textit{Median}} & \textit{Avg \& SD} & \textcolor{blue}{\textit{Median}} \\ \midrule
\textit{Seq2Seq}               & 0.20$\pm$1.22   & 0.00 & 0.33$\pm$1.58   & 0.22 & 0.37$\pm$1.67   & 0.22 \\
\textit{CodeBERT}              & 0.14$\pm$1.12   & 0.00 & 0.29$\pm$1.57   & 0.20 & 0.33$\pm$1.68   & 0.21 \\
\textit{CodeT5+}               & 0.11$\pm$0.97   & 0.00 & 0.22$\pm$1.36   & 0.08 & 0.27$\pm$1.51   & 0.20 \\
\textit{PLBart}                & 0.21$\pm$1.56   & 0.00 & 0.35$\pm$1.99   & 0.19 & 0.46$\pm$2.30   & 0.21 \\
\textit{ChatGPT}    & {0.32$\pm$1.51} & 0.00 & {0.67$\pm$2.13} & 0.28 & {0.70$\pm$2.17} & 0.29 \\ \midrule
\textit{All Models} & {0.20$\pm$1.28} & 0.00 & {0.37$\pm$1.73} & 0.21 & {0.43$\pm$1.87} & 0.22 \\
\bottomrule
\end{tabular}
\end{table}

Another aspect that influences the total computational cost required for the analysis is the type of operation performed by the code snippet. For instance, while logical operations (e.g., \texttt{and}, \texttt{xor}, \texttt{not}) and instructions that handle register contents (e.g., \texttt{inc}, \texttt{dec}, \texttt{mov}) are fast computed (i.e., $\sim0.25$s), instructions used to iterate over a variable, to compare two registers, or to perform conditional jumps (e.g., \texttt{cmp}, \texttt{loop}, \texttt{jns}) are less time-efficient. 
This is because arithmetical and logical operations are often simpler to implement because they involve basic bit-level manipulation. Contrarily, comparisons usually involve comparing values from different registers or memory locations, and conditional jumps depend on the result of these comparisons. This complexity can lead to longer execution times compared to simple logical operations.
\tablename~\ref{tab:examples_times} presents two examples of outliers in the computational cost analysis, \approach{}'s result, and the time needed for their evaluation.

\begin{table}[t]
\centering
\caption{Examples of code snippets with a high computational cost.}
\label{tab:examples_times}
\footnotesize
\begin{tabular}{
>{\centering\arraybackslash}m{4cm}
>{\centering\arraybackslash}m{4cm}
>{\centering\arraybackslash}m{1.5cm}
>{\centering\arraybackslash}m{1.5cm}}
\toprule
\textbf{Ground Truth} & \textbf{Model's Prediction} & \textbf{\approach{}} & \textbf{Time (s)} \\ 
\midrule
\texttt{cmp BYTE al, 2} \textbackslash n \texttt{je do\_inject} \textbackslash n \texttt{jmp while} & 	\texttt{cmp al, 2} \textbackslash n \texttt{jne while} \textbackslash n \texttt{jmp do\_inject} & 0 & 23.92 \\
\midrule
\texttt{cmp ax, bx} \textbackslash n \texttt{jne l3} \textbackslash n \texttt{jmp while}	& \texttt{cmp ax, bx} \textbackslash n \texttt{jne while} & 0 & 11.98 \\
\bottomrule
\end{tabular}
\end{table}

Both ground truth snippets perform similar operations: a comparison between two registers or a numerical value, a conditional jump based on the previous result, and an unconditional jump to a specific label. While in row \# 1 the code generated by the model has the same code complexity, in the second one the prediction exhibits lower complexity since the last jump is missing (i.e., \texttt{jmp while}).
Both predictions are classified as incorrect by \approach{} and take $\sim24$s and $\sim12$s, respectively.

While these outliers do increase the computational cost of our method, they represent a minimal fraction of all the generated samples, as in $17$ out of $2950$ predictions, i.e., only the 0.58\%. Specifically, there are $3$, $2$, $1$, $4$ and $7$ outliers for Seq2Seq, CodeBERT, CodeT5+, PLBart, and ChatGPT, respectively. The inclusion in our analyses of the outliers underscores the rigorousness of our methodology and highlights areas for future optimization in ACCA's execution efficiency.

Finally, to provide a context for the evaluation, we compared the computational cost of \approach{} with the ones of the output similarity metrics, which are automatic and time-saving solutions, and the ChatGPT-based assessment solutions. 
Unsurprisingly, we found that the output similarity metrics provide an average estimate of similarity in a very limited amount of time ($\sim0.004$ seconds on average per snippet), ranging from $0.001$ seconds for the exact match to $\sim0.01$ seconds for the SacreBLEU metric.
ChatGPT is also time-efficient, needing only $\sim0.003$ seconds to evaluate the correctness of the generated code with respect to the code description (ChatGPT-NL) and $\sim0.001$ for the comparison between predicted and ground truth snippets (ChatGPT-GT).
As a result, the computational costs of \approach{}, are higher than one of the baseline solutions since it depends on the non-negligible time needed by the binary execution. 

However, it is important to stress again that the output similarity metrics provide only an estimate of the code similarity rather than evaluating the code's correctness. Moreover, although ChatGPT provides limited computational time for the assessment, it is not an automated solution as it requires a non-trivial manual effort, including detailed instructions and several iterations with the human operator. On the contrary, our method is fully automated as it does not require any human intervention for the assessment.

Finally, it is worth noticing that the computational times of \approach{} are definitely lower than the average time required by human analysts to manually inspect the code, based on our experience. Indeed, since the human analyst needs to check both the NL description and the code snippet predicted by the models, in our experiments, the assessment of the semantic correctness required $\sim 20$ seconds on average per code snippet.

\section{Threats to Validity}
\label{sec:threats}
\noindent
\textbf{Relevance to Security.} \approach{} is designed as a tool to evaluate the semantic correctness of assembly code generated by AI models. Its primary function is to determine if AI-generated code behaves as intended when compared to a reference implementation. This involves assessing whether the generated code and the reference lead to the same final system state under the same initial conditions, thus ensuring semantic equivalence. While \approach{} is applicable in various contexts where assembly code generation is involved, our study specifically targets security-oriented code due to the critical importance of correctness in this domain. The choice to focus on security-oriented code, exemplified by using the extended version of the Shellcode\_IA32 dataset, stems from the challenges associated with generating assembly code for security purposes, e.g., exploit development.
It is crucial to remark that \approach{} does not perform a security assessment of the code itself; rather, it assesses the code's semantic correctness within security contexts. The tool does not identify vulnerabilities, exploits, or other security flaws in the generated code. Instead, it is a novel solution to automatically assess the semantics correctness of the code generated by models.

\noindent
\textbf{Relevance to the oracle problem in SE.} We acknowledge that the Oracle problem is a central challenge in software engineering, particularly in the domain of automated testing~\cite{DBLP:journals/jss/GiamatteiGPR24} and code generation~\cite{dinella2022toga,hossain2023neural}. The difficulty in establishing a definitive oracle that can accurately predict the correct output for any given input underlies many of the challenges in assessing the semantic correctness of AI-generated code. Our work indirectly addresses this problem by assessing semantic equivalence between AI-generated assembly code and a reference implementation. While we do not claim to solve the oracle problem, our approach provides a practical methodology for evaluating whether AI-generated code meets the intended specifications as represented by the reference code. This is especially relevant in contexts where the correctness of the code is critical, such as security-oriented applications. 

\noindent
\textbf{Automation Claims and Reference Implementation.}
\approach{}'s automation is predicated on the availability of a reference implementation, a common prerequisite in the field for assessing AI code generators. The reference implementations used for evaluation (i.e., the test set) are part of the same dataset that is used for training and validation purposes. Indeed, it is common practice in the field to randomly split the fine-tuning dataset into training, validation, and test sets, ensuring that the models are exposed to a diverse range of examples during the learning process. Hence, the reference for assessment is always available as long as there is a dataset to fine-tune models. 
In this work, we operate under the assumption that the reference implementation used for evaluation is reliable and accurately reflects the intended functionality as specified in the natural language descriptions. This assumption is consistent with standard practices in the field of code generation and evaluation, where output similarity metrics and other assessment methods rely on a trusted reference to determine the correctness of generated code. If we were to question the reliability of the reference implementations used for evaluation, it would logically extend to questioning the integrity of the entire dataset, including the portions used for training models. Such a scenario would imply that the models have been fine-tuned based on incorrect data, fundamentally challenging the premise of our study. However, this is beyond the scope of our work, which focuses on automating the assessment of the semantic correctness of AI-generated assembly code.

\noindent
\textbf{Assessment of AI code generators.} Our tool is deterministic in its assessment process. The evaluation criteria and procedures ACCA employs are consistent across all evaluations, ensuring that any given code snippet's assessment outcome remains the same under identical conditions. Hence, performing multiple training of the models, given that the models’ training is inherently stochastic, would only showcase the variability in AI model outputs, but this is not the focus of our work.
The benefit of executing multiple models’ training would be limited to a more comprehensive set of predictions to use for assessing the tool’s performance. However, we remark that our work performs an extensive evaluation of \approach{} by using the code generated by 5 distinct state-of-the-art models, for a total of $2950$ predictions to evaluate (i.e., $590$, the number of snippets of the test set, multiplied by the number of models, $5$), of which $1369$ are unique paris reference-predictions. This diverse set of models ensures that our evaluation encompasses a broad spectrum of AI-generated code, thereby enhancing the generalizability of our results. 
Finally, we highlight the computational resources to run models and the considerable time to perform manual analysis on all code snippets generated by the models, which would make our experiments prohibitive with multiple repetitions.

\noindent
\textbf{Dataset:} This work targets the assessment of AI-based solutions in the generation of offensive code for software exploits. The dataset used for our experiments fits perfectly with the scope of this work since it is the largest collection of offensive code available to date for code generation. This manually curated dataset contains high-quality and detailed descriptions of code, that are often not available in larger corpora for code generation. Indeed, the dataset provides NL descriptions both at the block and statement levels that are closer to the descriptions needed by the models for complex programming tasks.

\noindent
\textbf{Generalization and Offensive Code Generation}:
Although offensive code is different from general-purpose ones in terms of programming languages and characteristics, the proposed method can be applied to different assembly code generation scenarios. 
The decision to focus on offensive code generation is driven by the critical importance of code correctness assessment in this specific field. The use of models to generate offensive code, a research topic that is gaining increasing interest in software security~\cite{yang2023exploitgen,botacin2023gpthreats,natella2024ai}, poses unique challenges, requiring code generated by models to be properly evaluated. By focusing on exploits, the study addresses a specific and high-impact application of code generation, providing insights into the challenges of assessing the correctness of models in scenarios with stringent requirements.

\section{Related Work}
\label{sec:related}
\noindent
\textbf{Automatic Program Evaluation.}
Traditionally, the problem of automatic program assessment has been largely addressed for educational purposes, aiming to assist educators in the student work evaluation process. Insa and Silva~\cite{insa2018automatic,insa2015semi} presented a tool to assess Java programs by automatically validating different properties, such as the use of interfaces and class hierarchy. 
Romli \textit{et al.}~\cite{6985993} developed \textit{FaSt-Gen}, a framework of test data generation to cover both the functional and structural testing of programs for automatic assessment. 
Li \textit{et al.}~\cite{10.1145/2889160.2889204} leveraged random testing and dynamic symbolic execution (DSE), i.e., a software testing technique that simulates the execution of a program by providing symbolic inputs instead of concrete values. They generated test inputs and ran programs on these test inputs to compute values of behavioral similarity. Arifi \textit{et al.}~\cite{7474648} proposed a method to grade C programs in an educational context automatically. They measured the similarity between programs by comparing the outputs of their symbolic execution.
\textit{CASM-VERIFY}~\cite{lim2019automatic} is a tool to automatically check the equivalence of optimized assembly implementations of cryptographic algorithms. The tool decomposes the equivalence checking problem into several small sub-problems using a combination of concrete and symbolic evaluation. 
The use of symbolic execution to evaluate code similarity has been explored also for security applications. Luo \textit{et al.}~\cite{7823022} introduced a binary code similarity comparison method for code theft detection. Gao \textit{et al.}~\cite{gao2008binhunt} presented \textit{BinHunt}, a method to identify the semantic differences between an executable and its patched version, revealing the vulnerability that the patch eliminates.
Scalabrino \textit{et al.} focused on automatically assessing the understandability of code snippets by combining 121 existing and new metrics, including code-related, documentation-related, and developer-related metrics. They concluded, however, that these metrics are not suited to capture the complexity of code in practical applications.
Ullah and Oh~\cite{9470904} proposed a neural network-based solution to perform \textit{binary diffing} on x86 architecture binaries, i.e., the process of discovering the differences and similarities in functionality between two binary programs.
Leveraging symbolic execution to check semantic equivalence has been proposed and used in the area of compiler validation since compilers should preserve semantics. For example, Bera \textit{et al.}~\cite{bera2016practical} applied symbolic execution on the bytecode produced by the compilation with optimizations and that produced by the compilation without optimizations, in order to detect compiler bugs. Hawblitzel \textit{et al.}~\cite{hawblitzel2013will} detected compiler bugs by comparing assembly language outputs through symbolic execution. 
These solutions, however, require entire programs as input and do not work on portions of code (i.e., code snippets), which is often the case with AI-generated code since NMT is still far from generating entire complex functions, particularly in the context of offensive security.

\vspace{0.1cm}
\noindent
\textbf{Programming Language Code-oriented Metrics.}
In addition to state-of-the-art textual similarity metrics used as a baseline for the evaluation (see \S{}~\ref{subsec:metrics}), recent work proposed a set of novel code-oriented metrics, i.e., metrics created ad-hoc for specific programming languages, to automatically assess the correctness of the generated code.
Examples of code-oriented metrics are CodeBLEU~\cite{DBLP:journals/corr/abs-2009-10297} and RUBY~\cite{DBLP:conf/iwpc/TranTNNN19}, which were introduced to evaluate programs written in Java and C\#. 
However, these solutions rely on deeper program analysis, including syntax and dataflow match, and require compilable code to function, which prevents them from being language-agnostic. Indeed, none of the available code-oriented metrics is designed for low-level programming languages such as assembly.
Previous work on code generation also resorted to \textit{functional correctness} to evaluate the quality of the generated programs, where a code sample is considered correct if it passes a set of unit tests. 
Kulal \textit{et al.}~\cite{kulal2019spoc} used an evaluation metric based on functional correctness to address the problem of producing correct code starting from pseudocode. They generated \textit{k} code samples per problem and assessed the ratio of problems in which any of the \textit{k} samples passed the set of unit tests.
Chen \textit{et al}~\cite{chen2021evaluating} proposed \texttt{pass@k}, an unbiased and numerically stable implementation of this metric. They generated $n \geq k$ samples per task ($n = 200$ and $k \leq 100$), counted the number of correct samples $c \leq n$ that pass unit tests, and calculated an unbiased estimator to benchmark their models in the generation of Python programs from docstrings.
To estimate the functional correctness of a program, however, a set of unit tests needs to be manually constructed. This requires a significant effort that is often unfeasible for large amounts of generated code.

\vspace{0.1cm}
\noindent
\textbf{AI Generative for Security.}
Automatic exploit generation (AEG) research challenge consists of automatically generating working exploits~\cite{avgerinos2014automatic}. 
This task requires technical skills and expertise in low-level languages to gain full control of the memory layout and CPU registers and attack low-level mechanisms (e.g., heap metadata and stack return addresses). 
Given their recent advances, AI-code generators have become a new and attractive solution to help developers and security testers in this challenging task. 
Liguori \textit{et al.}~\cite{liguori2022can} released a dataset containing NL descriptions and assembly code extracted from software exploits. The authors performed an empirical analysis showing that NMT models can correctly generate assembly code snippets from NL and that in many cases can generate entire exploits with no errors. 
The authors extended the analysis to the generation of Python security-oriented code used to obfuscate software exploits from systems' protection mechanisms~\cite{liguori2021evil}.
Yang \textit{et al.}~\cite{yang2022dualsc} proposed a data-driven approach to software exploit generation and summarization as a dual learning problem. The approach exploits the symmetric structure between the two tasks via dual learning and uses a shallow Transformer model to learn them simultaneously.
Yang \textit{et al.}~\cite{yang2023exploitgen} proposed a novel template-augmented exploit code generation approach. The approach uses a rule-based template parser to generate augmented NL descriptions and uses a semantic attention layer to extract and calculate each layer’s representational information. The authors show that the proposed approach outperforms the state-of-the-art baselines from the previous studies of automatic code generation. 
Ruan \textit{et al.}~\cite{ruan2023prompt} designed an approach for software exploit generation based on prompt tuning. The solution aids the generation process by inserting trainable prompt tokens into the original input to simulate the pre-training stage of the model to take advantage of its prior knowledge distribution.
Xu \textit{et al.}~\cite{xu2023autopwn} introduced an artifact-assisted AEG solution that automatically summarizes the exploit patterns from artifacts of known exploits and uses them to guide the generation of new exploits. The authors implemented AutoPwn, an AEG system that automates the generation of heap exploits for Capture-The-Flag \textit{pwn} competitions. 
Recent work also explored the role of GPT-based models, including ChatGPT and Auto-GPT, in the offensive security domain. Botacin~\cite{botacin2023gpthreats} found that, by using these models, attackers can both create and deobfuscate malware by splitting the implementation of malicious behaviors into smaller building blocks.
Pa \textit{et al.}~\cite{pa2023attacker} and \cite{gupta2023chatgpt} proved the feasibility of generating malware and attack tools through the use of reverse psychology and \textit{jailbreak prompts}, i.e., maliciously crafted prompts able to bypass the ethical and privacy safeguards for abuse prevention of AI code generators like ChatGPT.
Gupta \textit{et al.}~\cite{gupta2023chatgpt} also examined the use of AI code generators to improve security measures, including cyber defense automation, reporting, threat intelligence, secure code generation and detection, attack identification, and malware detection.
Natella \textit{et al.}~\cite{natella2024ai} built a security-oriented evaluation benchmark to discuss potential use cases of AI code generators for offensive security. These use cases encompass attack surface analysis, OSINT, vulnerability exploitation, and post exploitation activities. The authors concluded that cybersecurity professionals must embrace AI code generators to prevent attacks more efficiently.
All previous work uses state-of-the-art output similarity metrics or performs manual analysis to assess the correctness of AI-generated code/programs. 

Our work is complementary to previous ones. Indeed, this work proposes a method that leverages symbolic execution to automatically assess the correctness of low-level code snippets used in security contexts. 
Since the method does not necessarily require full programs in inputs, it is suitable for assessing AI-generated code because they are often incomplete or non-compilable programs. 
Moreover, the proposed method does not require any human intervention, yet, differently from traditional text similarity metrics, which are commonly used to assess the performance of AI-generated code, its accuracy is comparable to human evaluation. 

\section{Conclusion}
\label{sec:conclusion}
In this paper, we addressed the automatic correctness of the code generated by AI code generators. We proposed a fully automated method, named \approach{}, that uses symbolic execution to assess the correctness of security-oriented code without any human effort.

We used our method to evaluate the performance of four state-of-the-art code generators in the generation of offensive assembly from NL descriptions and compared the results with the human evaluation and different baseline solutions, including state-of-the-art output similarity metrics and the well-known ChatGPT.

Our experiments showed that \approach{} provides results almost equal and is the most correlated assessment solution to human evaluation, which is considered the golden standard in the field.
Moreover, the analysis of the computational cost revealed that the time to assess every code snippet is $\sim 0.17$s on average, which is lower than the average time required by human analysts to manually inspect the code, based on our experience.

In future work, integrating \approach{} with advanced static code analysis and binary matching approaches could significantly enhance its assessment capabilities. For instance, static code analysis could provide additional insights into potential vulnerabilities or inefficiencies in the generated code~\cite{sui2016svf}, while binary matching could extend \approach{}'s applicability to compiled binaries~\cite{gui2022cross}, offering a more holistic view of code correctness and security.

Future work also includes supporting the automatic correctness assessment for other programming languages. We are actively exploring the extension of \approach{} to evaluate code in high-level programming languages such as Python. This expansion will address the growing need for semantic correctness assessment in diverse coding environments, further enhancing \approach{}'s applicability and impact in the assessment of code generation models.

\section*{Acknowledgements}
This work has been partially supported by the University of Naples Federico II in the frame of the MUR PRIN 2022 program, project FLEGREA, CUP E53D23007950001 (\url{https://flegrea.github.io}). 
We are grateful to our former student Emiliano Fiorenza for his help in the early stages of this work.



\bibliographystyle{elsarticle-num} 
\bibliography{biblio}





\end{document}